%% file: main.tex
\documentclass[10pt,conference]{IEEEtran}
\IEEEoverridecommandlockouts
\usepackage{cite}
\usepackage{amsmath,amssymb,amsfonts}
\usepackage{algorithmic}
\usepackage{graphicx}
\usepackage{textcomp}
\usepackage{xcolor, colortbl}
\usepackage{booktabs}
\usepackage{multirow}
\usepackage{xspace}
\usepackage{caption}
\usepackage{subcaption}
\usepackage{footnote}
\usepackage{url}
\usepackage{tcolorbox}
\usepackage{booktabs}

\usepackage{listings}
\usepackage{balance}

\definecolor{Gray}{gray}{0.9}

\def\BibTeX{{\rm B\kern-.05em{\sc i\kern-.025em b}\kern-.08em
    T\kern-.1667em\lower.7ex\hbox{E}\kern-.125emX}}
    
\newcommand\revised[1]{\textcolor{black}{#1}}

\begin{document}

\title{An Empirical Study on the Usage of\\BERT Models for Code Completion}

\author{
\IEEEauthorblockN{Matteo Ciniselli\IEEEauthorrefmark{1}, Nathan Cooper\IEEEauthorrefmark{2}, Luca Pascarella\IEEEauthorrefmark{1}, Denys Poshyvanyk\IEEEauthorrefmark{2},\\Massimiliano Di Penta\IEEEauthorrefmark{3}, Gabriele Bavota\IEEEauthorrefmark{1}}

\IEEEauthorblockA{\IEEEauthorrefmark{1}\textit{SEART @ Software Institute, Universit\`{a} della Svizzera italiana (USI), Switzerland}}
\IEEEauthorblockA{\IEEEauthorrefmark{2}\textit{SEMERU @ Computer Science Department, William and Mary, USA}}
\IEEEauthorblockA{\IEEEauthorrefmark{3}\textit{Department of Engineering, University of Sannio, Italy}}
}
\maketitle

\newcommand{\roberta}{{RoBERTa}\xspace}
\newcommand{\mask}{{\tt <MASK>}\xspace}
\newcommand{\ie}{\emph{i.e.,}\xspace}
\newcommand{\eg}{\emph{e.g.,}\xspace}
\newcommand{\etc}{etc.\xspace}
\newcommand{\etal}{\emph{et~al.}\xspace}
\newcommand{\secref}[1]{Section~\ref{#1}\xspace}
\newcommand{\figref}[1]{Fig.~\ref{#1}\xspace}
\newcommand{\listref}[1]{Listing~\ref{#1}\xspace}
\newcommand{\tabref}[1]{Table~\ref{#1}\xspace}
\newcommand{\tool}[1]{{\sc #1}\xspace}

\newcommand{\totalTokens}{xxx,xxx\xspace}
\newcommand{\totalMethods}{xx\xspace}
\newcommand{\totalRepo}{xx\xspace}

\newboolean{showcomments}

\setboolean{showcomments}{true}

\ifthenelse{\boolean{showcomments}}
  {\newcommand{\nb}[2]{
    \fbox{\bfseries\sffamily\scriptsize#1}
    {\sf\small$\blacktriangleright$\textit{#2}$\blacktriangleleft$}
   }
  }
  {\newcommand{\nb}[2]{}
  }

\newcommand\MATTEO[1]{\textcolor{red}{\nb{MATTEO}{#1}}}
\newcommand\NATHAN[1]{\textcolor{red}{\nb{NATHAN}{#1}}}
\newcommand\LUCA[1]{\textcolor{red}{\nb{LUCA}{#1}}}
\newcommand\DENYS[1]{\textcolor{red}{\nb{DENYS}{#1}}}
\newcommand\MAX[1]{\textcolor{red}{\nb{MAX}{#1}}}
\newcommand\GABRIELE[1]{\textcolor{red}{\nb{GABRIELE}{#1}}}

\pagenumbering{gobble}

\begin{abstract}
Code completion is one of the main features of modern Integrated Development Environments (IDEs). Its objective is to speed up code writing by predicting the next code token(s) the developer is likely to write. Research in this area has substantially bolstered the predictive performance of these techniques. However, the support to developers is still limited to the prediction of the next few tokens to type. In this work, we take a step further in this direction by presenting a large-scale empirical study aimed at exploring the capabilities of state-of-the-art deep learning (DL) models in supporting code completion at different granularity levels, including single tokens, one or multiple entire statements, up to entire code blocks (\eg the iterated block of a \textit{for} loop). To this aim, we train and test several adapted variants of the recently proposed \roberta model, and evaluate its predictions from several perspectives, including: (i) metrics usually adopted when assessing DL generative models (\ie BLEU score and Levenshtein distance); (ii) the percentage of perfect predictions (\ie the predicted code snippets that match those written by developers); and (iii) the ``semantic'' equivalence of the generated code as compared to the one written by developers. The achieved results show that BERT models represent a viable solution for code completion, with perfect predictions ranging from $\sim$7\%, obtained when asking the model to guess entire blocks, up to $\sim$58\%, reached in the simpler scenario of few tokens masked from the same code statement. 
\end{abstract}

\thispagestyle{plain}

\input{introduction}
\input{approach}
\input{design}
\input{results}

\input{threats}
\input{related}

\input{conclusion}

\section*{Acknowledgment}
This project has received funding from the European Research Council (ERC) under the European Union's Horizon 2020 research and innovation programme (grant agreement No. 851720). W\&M team was supported in part by the NSF CCF-1955853, CCF-2007246 and CCF-1815186 grants. Any opinions, findings, and conclusions expressed herein are the authors' and do not necessarily reflect those of the sponsors. 

\newpage
\balance
\bibliography{main}
\bibliographystyle{IEEEtran}

\end{document}

%% file: introduction.tex
\section{Introduction} \label{sec:intro}
The software development landscape is continuously changing, with new and evolving programming languages, frameworks, and APIs. This makes writing code by heart quite challenging even for the most experienced developers. For this reason, code completion is considered to be one of the ``killer'' features of modern Integrated Development Environments (IDEs)~\cite{Bruch:fse2009,Robb2010a,kim2020code}: It can provide developers with recommendations about the next code token (\eg a method call) to write given the code already written in the IDE, thus speeding up software development and preventing mistakes~\cite{han2009code,han2011code}.

The existing literature has documented major advances of code completion tools, with their recommendations ranging from mere alphabetical lists of the next token to write given the characters already typed (\eg a list of possible method calls matching the first character typed by the developer) to ``intelligent'' completions considering the context surrounding the code \cite{Bruch:fse2009,Robb2010a}, the history of code changes \cite{Robb2010a}, and/or coding patterns mined from software repositories \cite{Hindle:icse2012,Nguyen:icse2012,Tu:fse2014,Asaduzzaman2014,Nguyen:msr2016,niu2017api,Hellendoorn:fse2017}. 

Last, but not least, Deep Learning (DL)  models have been applied to code completion \cite{Karampatsis:DLareBest,kim2020code,alon2019structural,svyatkovskiy2020intellicode,White2015}, setting new standards in terms of prediction performance. Although the performance of code completion techniques substantially improved over time, the type of support they provide to developers has not evolved at the same pace, and are mostly only capable of predicting a single token. Only a few recent studies focus on predicting multiple contiguous tokens \cite{alon2019structural,svyatkovskiy2020intellicode}. 

We present a large-scale empirical study exploring the limits and capabilities of state-of-the-art DL models to support code completion. Besides generating the next token(s) the developer is likely to write, we apply DL models to the generation of entire statements and code blocks (\eg the body of an {\tt if} statement). Among the many DL models proposed in the literature, we decided to adapt the \roberta model recently proposed by Liu \etal \cite{roberta}. \roberta is a BERT (Bidirectional Encoder Representations from Transformers) model \revised{\cite{Delvin:2019}} using a pre-training task in which random words in the input sentences are masked out using a special \mask token, with the model in charge of predicting the masked words. The \roberta pre-training task formulation is particularly suited for code completion: The input sentences can be seen as code statements and the masked words as masked code tokens. Also, as compared to statistical language models, BERT models have the advantage of considering both the words preceding and following the masked words to perform the prediction.

One limitation of the \roberta pre-training task is that $n$ \mask tokens must be used to mask $n$ code tokens, thus implicitly suggesting to the model how many code tokens must be generated to autocomplete the masked statement. This would not be realistic in a real usage scenario, in which the code completion engine must \emph{guess} the tokens to generate, without the developer suggesting how many tokens must be generated. For this reason, we adapted the \roberta pre-training objective to be able to guess, from a single \mask token masking one or more code tokens in the given statements, which and how many code tokens must be generated.

\revised{Also, note that the goal of our study is not to show that \roberta is the best option for neural-based code completion. Our work focuses on empirically exploring the capabilities of DL-based code completion and \roberta has been chosen as representative of the state-of-the-art DL techniques.}

\newpage

We train three different variants of the \roberta model, specialized in: (i) \emph{token-level} predictions, namely classic code completion in which the model is used to guess the last $n$ tokens in a statement the developer started writing; (ii) \emph{construct-level} predictions, in which the model is used to predict specific code constructs (\eg the condition of an {\tt if} statement) that can be particularly useful to developers while writing code; and (iii) \emph{block-level} predictions, with the masked code spanning one or more entire statements composing a code block (\eg the iterated block of a {\tt for} loop).

We analyze the quality of the generated predictions from several different perspectives, including: (i) metrics usually adopted in the literature when assessing DL generative models (\ie BLEU score \cite{dreyer:bleu} and Levenshtein distance \cite{levenshtein1966}); (ii) the percentage of perfect predictions (\ie the predicted code is exactly the same as the one written by the developer); and (iii) the ``semantic'' equivalence of non-perfect predictions, namely cases in which the code predicted by the model is different as compared to the one originally written by the developers but equivalent in terms of provided functionality. We also compare the \roberta model with a state-of-the-art \textit{n}-gram model presented by Hellendoorn and Devanbu~\cite{Hellendoorn:fse2017}.

The achieved results show that, for the typical code completion task (\ie  \emph{token-level}), \roberta is able to correctly guess all masked tokens in 39\% to 58\% of cases, depending on the specific dataset and code representation we use. When the code completion task concerns more challenging scenarios such as the \emph{construct-level} predictions, the performance drops by 10\%-15\%. Finally, in the most challenging scenario in which we mask entire blocks, \roberta shows its limitations, being able to correctly reconstruct the masked block only in 7\%-9\% of the cases and, in particular, when the masked block is quite short in terms of tokens. When compared to the \textit{n}-gram model~\cite{Hellendoorn:fse2017}, the performance of \roberta is substantially better. However, this gain of performance comes at a much higher training cost.


%% file: approach.tex
\section{Using \roberta for Feature Completion} \label{sec:approach}

\begin{figure}[h]
	\vspace{-0.1cm}
	\centering
	\includegraphics[width=\linewidth]{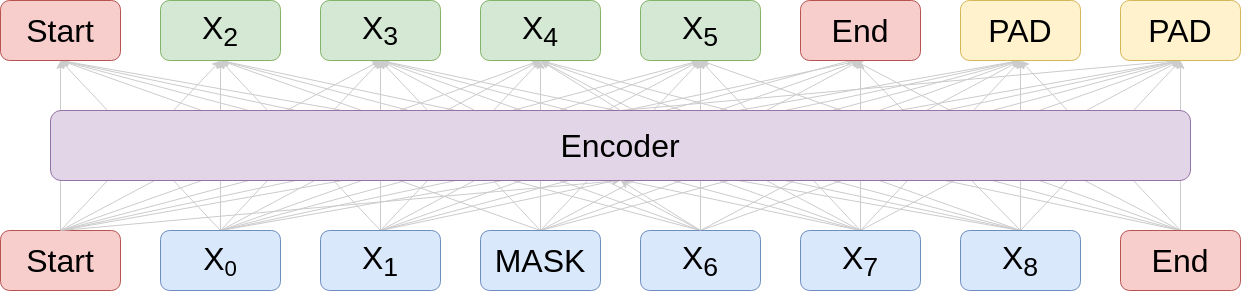}
	\caption{Modified pretraining task for extending BERT to predict a span of tokens from a single masked token.}
	\label{fig:span_pretraining}
\end{figure}

Our study leverages an off-the-shelf \roberta model, which is an Encoder-Transformer architecture. Details about the \roberta model are provided in a report by Liu \etal \cite{roberta}, while we focus on explaining why it represents a suitable choice for code completion. BERT-based models, such as \roberta, use a special pretraining where random words in the input sentence are masked out with a special \mask token. This pretraining task is very well-suited to simulate a code completion task, in which the input is an incomplete code snippet the developer is writing and the masked tokens represent the code needed to autocomplete the snippet. 
However, one limitation of such a pretraining is that when attempting to predict multiple tokens, such as an entire masked \textit{if} condition, it requires the number of tokens to generate to be known, due to the fixed sequence length of Transformers \cite{attention}. To overcome this issue, we propose to use a version of the T5 \cite{t5} pretraining objective, in which spans of tokens are masked using a single token. In \figref{fig:span_pretraining} four tokens (\ie $X_2$ to $X_4$) are masked with a single \mask token.

As detailed in \secref{sec:design}, we trained twelve \roberta models dealing with different code completion tasks. \revised{It is important to clarify the choice of training different models in the context of a BERT model. In the original work presenting BERT \cite{Delvin:2019}, two characteristics of this family of models are presented: (i) the ability to take advantage of a pre-training task to boost performance; and (ii) the possibility to fine-tune a pre-trained model on several tasks, taking advantage of transfer learning. To make a concrete example, it is possible to pre-train a BERT model on a corpus of English text (\eg Wikipedia pages) by randomly masking some words in the text and asking the model to predict the masked words. Such a pre-training task is self-supervised, and can scale to large datasets since it does not require any manual labeling. Once the model is pre-trained, it can be specialized (fine-tuned) to support different tasks such as language translation, sentiment identification, \etc The result will be a single model able to support different tasks and, possibly, taking advantage of what it learned for a specific task (\eg language translation) to also improve its performance in a different task (\eg sentiment identification). In our study, we do not leverage a pre-training of \roberta, because the fine-tuning tasks (\ie predict specific parts of code we mask) would be very similar to what is considered a pre-training (\ie predict masked tokens) in BERT models. The only difference, as said, is that we mask multiple tokens with a single \mask token. Also, since this is an exploratory empirical study in which we want to assess the performance of \roberta models in simple and more challenging code completion scenarios and datasets, we decided to train different models for each code completion task, avoiding the possible confounding factor introduced by transfer learning. Indeed, as shown in recent work \cite{Arivazhagan:2019} fine-tuning a model for several tasks can have a substantial influence on the model's performance.
}

To train the models, we used the Python \emph{transformers}~\cite{Wolf2019HuggingFacesTS} library. Besides training the \roberta models, we also trained a tokenizer for each of them. We trained a Byte Pair Encoding (BPE) \cite{bpe} model using the HuggingFace's \emph{tokenizers}~\cite{tokenizers} Python library. BPE uses bytes as vocabulary, allowing it to tokenize every text without requiring the unknown token often used in applications of DL to NLP, helping to solve the out-of-vocabulary problem for code~\cite{karampatsis2020big}.

%
%
%
%

%% file: design.tex
\section{Study Design} \label{sec:design}

The study \emph{goal} is to assess the effectiveness of \roberta in predicting masked code tokens at different granularity levels. We address the following research questions (RQs):

{\bf RQ$_1$:} \emph{Are BERT models a viable approach to learn how to autocomplete code?} This RQ investigates the extent to which \roberta can be used to predict missing code tokens. We assess the quality of the generated predictions from both a quantitative (\ie BLEU score, Levenshtein distance) and a qualitative (\ie perfect predictions, potential usefulness of wrong predictions) perspective.   RQ$_1$ is further detailed in the following three sub-RQs:

{\bf RQ$_{1.1}$:} \emph{To what extent does the number of masked tokens impact the prediction quality?} We train and test \roberta on datasets in which masked code tokens span from few contiguous tokens in a given statement to multiple missing statements composing a code block. RQ$_{1.1}$ explores the limits of BERT models when considering simple and more challenging code completion scenarios.

{\bf RQ$_{1.2}$:} \emph{To what extent reducing the vocabulary by abstracting the source code helps in the prediction task?} Previous applications of DL to source code applied code abstraction \cite{Tufano:icse2019,Tufano:tosem2019} to avoid the open vocabulary problem typical of source code. As said, \roberta does not suffer from this limitation thanks to the usage of BPE. However, we still want to investigate whether reducing the vocabulary further helps the learning.
In this research question, we test whether abstracting the code using the approach proposed in the literature \cite{Tufano:icse2019,Tufano:tosem2019} helps the model learning. We compare the prediction performances with and without applying abstraction, 
while abstraction is not required when using BPE. 

{\bf RQ$_{1.3}$:} \emph{To what extent are the performance of the model influenced by the specificity of the dataset employed for training and testing it?} While it is reasonable to expect that larger training datasets tend to help deep learning models, we are interested in answering RQ$_{1.3}$ from a different perspective. To address this RQ, we compare the autocompletion performances on two different datasets: a first, more general one, composed of Java methods; and a second, more specific one, composed of methods from Android apps. While the programming language is the same, the second dataset makes heavy use of Android APIs, and it is likely that the same APIs are used for similar purposes, \eg app features dealing with GPS positioning share common API usages. We expect this to create ``regularities'' in the Android dataset to help model learning.

{\bf RQ$_2$:} \emph{How does the \roberta model compare to a state-of-the-art n-gram model?} An alternative to DL models is represented by statistical language models based on \textit{n}-grams. In this research question, we compare the trained models to the state-of-the-art \textit{n}-gram cached model \cite{Hellendoorn:fse2017}.

\subsection{Dataset Construction}
To create the \emph{Java dataset}, we started from the CodeSearchNet Java Dataset provided by Husain \etal \cite{Java:CodeSearchNet}.  This dataset contains over 1.5M Java methods collected from open-source, non-fork, GitHub repositories. For details on how the dataset has been built, see a report by Husein \etal \cite{Java:CodeSearchNet}. For our work, the most important criteria used in the dataset building are: (i) excluding methods of fewer than three lines; (ii) removing near-duplicate methods using deduplication algorithm from CodeSearchNet \cite{Deduplication}; 
and (iii) removing methods with the name containing the ``test'' substring in an attempt to remove test methods; methods named ``toString'' are removed as well. 

To build the \emph{Android dataset} we adopted a similar procedure. We cloned from GitHub the set of 8,431 open-source Android apps available in the AndroidTimeMachine dataset \cite{Geiger:2018}. Then, we extracted from each project's latest snapshot the list of methods. This resulted in a total of $\sim$2.2M methods. Then, we applied the same filtering heuristics defined for the Java dataset, ending up with 654,224 methods.

Since one of the goals of our study is also to compare the performance of \roberta when applied on a more generic (Java) and a more specific (Android) dataset, we randomly selected from the Java dataset 654,224 methods, to match the size of the Android dataset. 

\subsubsection{Dataset Processing}
\label{sec:datasetProcessing}
We processed the Java and Android datasets to create several versions of them with the goal of answering our RQs. The created datasets are summarized in \tabref{tab:dataset} and described in the following.

Methods for which parsing errors occurred during the abstraction process were excluded from both datasets since we want to compare the performance of the BERT models when using/not using abstraction. Thus, the same instances should be included in the \emph{raw} and in the \emph{abstracted} datasets. This left the Java dataset with 634,799 methods, and the Android one with 532,096. For each of those methods, both the raw and the abstract versions are available. As a final step, we created three versions of each dataset, applying the following token masking procedures to both the raw and the abstract code.

\textbf{Token masking.} For each code line $l$ in each method having more than one token we mask its last $x$ tokens, where $x$ is a random number between $1$ $\dots$ $n-1$, where $n$ is the number of tokens composing $l$. 
The purpose of token-masking is to simulate a typical code completion scenario: A developer starts writing a code line, and the tool recommends how to complete it. Given a method $m$ having $k$ lines with more than one token, we generate $k$ versions of $m$, each of them having one and only one line with the last $x$ tokens masked. We set the maximum number of masked tokens to 10 (\ie if $x > 10$ then $x=10$).

\textbf{Construct masking.} We selected a number of code constructs for which it could be particularly useful to be supported with automated code completion. Given a method $m$, we use the srcML \cite{SrcML} toolkit to identify all $m$'s tokens used to: (i) define the complete condition of an {\tt if} statement or of a {\tt while}/{\tt for} loop (\eg in a statement having {\tt for(int i=0; i<data.size(); i++)} we identify all tokens between parenthesis as those used to define the {\tt for} loop); (ii) define the parameters in a method call (\eg in {\tt copyFile(source, target)} the tokens ``{\tt source}'', ``{\tt ,}'', and ``{\tt target}'' are identified); and (iii) define the exception caught in a {\tt catch} statement (\eg in {\tt catch(IOException io)} we identify {\tt IOException io} as the involved tokens). For $m$ this results in a set $S$=\{$T_1$, $T_2$, $\dots$, $T_n$\}, where $T_i$ represents a set of relevant tokens for one of the previously mentioned constructs (\eg $T_i$ is the set of tokens used to define the {\tt for } loop condition). 

Given $m$, we generate $|S|$ versions of it, each one having one of the subject constructs masked. Also in this case we set the maximum number of masked tokens to 10. This means that if a construct requires more than 10 tokens to be masked, it is not masked in our dataset.

\textbf{Block masking.} We use srcML to identify in each method $m$ its code blocks. We define a code block as the code enclosed between two curly brackets. For example, a block may be, besides the method body itself, the code executed in a {\tt for}/{\tt while} loop, when an {\tt if}/{\tt else}/{\tt else if} condition is satisfied, \etc Then, given $k$ the number of blocks identified in $m$, we create $k$ versions of $m$ each one having a specific code block masked. We set the maximum size of the masked block to two complete statements. This means that if a block is composed of more than two statements, it is not masked.

\input{tables/design-tables}

To address RQ$_{1.2}$, we used the {\tt src2abs} tool by Tufano \etal \cite{Tufano:icse2019,Tufano:tosem2019} to generate an abstract version of each dataset. This abstraction process has been proposed to reduce the source code vocabulary size, providing an expressive yet vocabulary-limited representation. For example, all variable names present in a method are abstracted as \texttt{VAR\_X}, where {\tt X} indicates the number of the variable in the method (\eg the first variable is abstracted as \texttt{VAR\_1}, the second as \texttt{VAR\_2}, \etc). All keywords of the language and punctuation symbols are left unchanged. Similarly, very frequent identifiers and literals (\ie \texttt{i}, \texttt{j}, \texttt{index}) are treated as \emph{idioms} and not abstracted. The abstraction process is described in \cite{Tufano:tosem2019}.

In summary, our study comprises twelve datasets: For each of the two domains (Java or Android) there are two code representations (raw or abstract), each of them with three different masking levels (token, construct, block).

\subsubsection{Creating Training, Evaluation, and Test sets}
Starting from the twelve datasets, we created the training, evaluation, and test sets in \tabref{tab:dataset}. \tabref{tab:dataset} only shows six datasets because of space limitations. However, the data for both the raw and abstract datasets is exactly the same in terms of the number of instances/tokens (only the code representation changes). 

As a first step, we filtered out specific instances from our datasets. First, when using generative deep learning models, the variability in length of the sentences (in our case, methods) provided as input can affect the training and performance of the model, even when techniques such as padding are employed. For this reason, we analyzed the distribution of methods length in our dataset, finding that two-thirds of them are composed of at most 100 tokens. For this reason, as done by Tufano \etal~\cite{Tufano:tosem2019}, we excluded from our datasets all methods having more than 100 tokens. Second, the \roberta model cannot efficiently handle cases in which the \emph{masked} tokens are more than the \emph{non-masked} tokens. This often happens, for example, when masking the entire method body in the block-level masking approach. Thus, those instances are excluded as well. Finally, we performed method de-duplication again, keeping for each group of duplicates one random instance only. While this was already done in the very first step of the dataset creation, it could happen that, after abstraction, two methods are equal even if their raw code is different (\eg they only differ for the value of a variable name that, however, is abstracted in both cases as $VAR_1$).  

After the filtering steps, we split each of the twelve datasets into training (80\%), evaluation (10\%), and test (10\%) sets. \revised{While the methods in the dataset are randomly ordered,} the splitting we performed was not random to avoid biasing the learning. To explain this point, let us consider the case of the \emph{block masking} dataset. Given a method $m$ having $k$ blocks in it, we add in the dataset $k$ versions of $m$, each having one and only one block masked. Suppose that $m$ contains two blocks $b_1$ and $b_2$, thus leading to two versions of $m$: One in which $b_1$ is masked ($m_{b_1}$) and $b_2$ is not and \emph{vice versa} ($m_{b_2}$). With a random splitting, it could happen that $m_{b_1}$ is assigned to the training set and $m_{b_2}$ to the test set. However, in $m_{b_1}$ the $b_2$ is not masked. Thus, when the model has to guess the tokens masked in $m_{b_2}$ it would have the solution in the training set, resulting in boosted prediction performance. For this reason, we take the first 80\% of the methods in each dataset and assign all of their masked versions to the training set. Then, we proceed in the same way with evaluation and test sets. 

Using this procedure, we obtained the datasets described in \tabref{tab:dataset}. Important to note is that, given the original size of the datasets using token-level and construct-level masking, we decided to cap the training set to 750k instances (no changes were done in the evaluation and test sets). This was necessary given the computationally-expensive process of training twelve different \roberta models (one for each dataset). Also, the size of the evaluation and test sets is slightly different since, as explained before, we split the dataset based on the methods (not on their masked versions) and each method can result in a different number of its generated masked versions.

\input{tables/hyperparameters}

\subsection{Data Collection and Analysis}
\label{sec:analysis}
After having obtained the twelve triplets of $\langle$training, evaluation, test$\rangle$ sets, we trained and tested twelve \roberta models using the best configuration we identified through a hyperparameter tuning procedure. 

We performed hyperparameter tuning using the Weights \& Biases's \cite{wandb} Python library on a Linux server with an Nvidia RTX Titan GPU. \tabref{tab:hyper} reports the hyperparameters we tuned, the range of values we tested for them, and the value in the best configuration we found. Besides those parameters, we used an attention dropout probability of 0.1, and an hidden layer dropout probability of 0.3. For the tokenizer, the vocabulary size was set to 50k for the raw datasets, and to the vocabulary size of the dataset for the abstract dataset. The hyperparameter search was performed using the training and the evaluation sets of the Android abstract dataset with raw masking. We picked as the best configuration the one that, when applied to the evaluation set, was able to obtain the highest number of ``perfect predictions''. We define as ``perfect'' a prediction that exactly matches the code written by the developers. Thus, the model correctly guesses \emph{all} masked tokens. If one of the masked tokens is different we do not consider the prediction as ``perfect''. While, in principle, a different hyperparameter tuning would be necessary for each dataset, such a process is extremely expensive, and preliminary investigations we performed on a subset of the other datasets showed minor differences in the achieved best configuration.

The training was performed across servers using their GPUs. The first was equipped with an Nvidia Tesla V100S, the second with an Nvidia RTX Titan, and the third with 3 Nvidia GTX 1080Ti. The training time strongly depends on the size of the dataset and the used server but ranged between 28 and 114 hours per model. Note that, once trained, each model can be used to perform predictions in the split of a second (on average, 0.12 second on a laptop CPU), thus making them a viable solution for ``real-time'' code completion.

We train each model for a maximum of 50 epochs. However, we adopted the following stopping condition. At the end of each training epoch, we executed the model on the evaluation set and we compute the number of perfect predictions. If we observe that, during the training, the performance of the model is worsening in terms of perfect predictions on the evaluation set (\ie the model is likely overfitting to the training set), we stop the training. In particular, given a model trained for $n^{th}$ epoch, we stop the training if the number of perfect predictions on the evaluation set is lower than the number of perfect predictions achieved after the $n-4$ epoch. This ensures that the models can have some fluctuations in performance for up to three epochs. Then, if it is still not improving, we stop its training and take the best model (in terms of perfect predictions on the evaluation test) obtained up to that moment. None of the models was trained for the whole 50 epochs.

By running each trained model on the corresponding test set we compute the following metrics:

\emph{The BLEU-n score \cite{dreyer:bleu}}. The BLEU score is a metric for assessing the quality of automatically translated text \cite{dreyer:bleu}. We use four variants of BLEU, namely BLEU-1, BLEU-2, BLEU-3, and BLEU-4. A BLEU-n variant computes the BLEU score by considering the n-grams in the generated text. Most of previous work in the SE literature adopt the BLEU-4 score  \cite{Gu:2016,Jiang:ASE'17,Watson:icse2020}. However, such a variant cannot be computed when the target prediction (in our case, the number of masked tokens) is lower than four. For this reason, we compute the four different versions: BLEU-1 can be computed for all predictions, while BLEU-n with n$>$1 only for predictions having a length (\ie number of tokens) higher or equal than $n$. The BLEU score ranges between 0\% and 100\%, with 100\% indicating, in our case, that the code generated for the masked tokens is identical to the reference one.

\emph{The Levenshtein distance \cite{levenshtein1966}}. To provide a proxy measure of the effort needed by developers in order to convert a prediction generated by the model into the reference (correct) code, we compute the Levenshtein distance at token-level: This can be defined as the minimum number of token edits (insertions, deletions or substitutions) needed to convert the predicted code into the reference one. Since such a measure is not normalized, it is difficult to interpret it in our context. Indeed, saying that five tokens must be changed to obtain the reference code says little without knowing the number of tokens in the reference code. For this reason, we normalize such a value by dividing it by the number of tokens in the longest sequence among the predicted and the reference code.

\emph{The percentage of perfect predictions}, which we statistically compare between abstract and raw datasets using the McNemar's test~\cite{mcnemar} and Odds Ratios (ORs). In this, as in other cases, we cope with multiple tests by adjusting p-values using the Benjamini-Hochberg correction~\cite{bh}. \smallskip  

We also analyze what happens in the case of non-perfect predictions. We manually analyzed a sample of non-perfect predictions to assess whether, while different, they were ``semantically equivalent'' to the original code written by developers. This could happen, for example, in case the masked code is {\tt return (x+1);} while the predicted code is {\tt return x+1;}. To do such an analysis we selected from each of the twelve test sets we have 100 non-perfect predictions, randomly picking 25 of them from each of four buckets: predictions having a Levenshtein distance between (i) n$>$0 and 0.24 (note that if n=0 this means that the prediction is perfect; (ii) 0.25 and 0.49; (iii) 0.50 and 0.74; and (iv) 0.75 and 1.00. Then, one of the authors inspected all 1,200 (100 $\times$ 12 datasets) instances, classifying each of them as ``semantically equivalent'' or not. Those classified as ``semantically equivalent'' have been double-checked by a second author to avoid subjectivity issues. We report, for each of the four Levenshtein distance intervals, the percentage of semantically equivalent predictions we found in each dataset. We statistically compare Levensthein distances and BLEU scores for the abstract dataset and the raw one using Wilcoxon signed-rank test, whereas we compare results for Android and Java using the Wilcoxon rank-sum test.

To address RQ$_2$, for all datasets, we compare the performance of \roberta with that of the state-of-the-art cached \textit{n}-gram model~\cite{Hellendoorn:fse2017} using the implementation made available by the authors \cite{ngram}. We tried to design a fair comparison, despite the fact that the \textit{n}-gram model is designed to predict a single token given the $n$ tokens preceding it. Thus, in a scenario in which we mask more than one token, we use the $n$-gram model in the following way: We run it to predict each masked token in isolation. Then, we join all predictions to generate the final string (\ie set of previously masked tokens). The $n$-gram models are trained on the same training sets used by \roberta without, however, masked tokens. We compare the two approaches in terms of perfect predictions generated on the test sets. A statistical comparison is performed using the McNemar's test \cite{mcnemar} and ORs.  

\subsection{Replication Package}
\label{sub:replication}
The datasets, the code implementing the \roberta models, and detailed results of statistical tests as well as the scripts used to run these tests are publicly available \cite{replication}.

%% file: tables/design-tables.tex


\begin{table}[ht]
	\centering
	\caption{Study datasets. One instance corresponds to a method with masked token(s).\vspace{-0.2cm}}
	\scriptsize
	\label{tab:dataset}
	\begin{tabular}{lllrr}
	\toprule
	\multirow{2}{*}{\bf Domain} & {\bf Masking} & \multirow{2}{*}{\bf Dataset} & \multirow{2}{*}{\bf \#Instances} & \multirow{2}{*}{\bf \#Tokens}\\ 
	& {\bf Level} & \\\midrule
	                      & \multirow{3}{*}{Token} & Training & 750k & 46.9M\\
	                      &                                & Evaluation & 215k & 13.4M\\
	                      &                                & Test & 219k & 13.6M\\
	                      \addlinespace[0.08cm]
	                      & \multirow{3}{*}{Construct} & Training & 750k & 48.2M\\
	               Java   &                                    & Evaluation & 104k & 6.7M\\
	                      &                                    & Test & 106k & 6.7M\\
	                      \addlinespace[0.08cm]
	                      & \multirow{3}{*}{Block} & Training & 298k &19.1M\\
	                      &                                & Evaluation & 39k & 2.5M\\
	                      &                                & Test & 40k & 2.5M\\
	                      \addlinespace[0.08cm]\hline\addlinespace[0.08cm]
	                      
	                         & \multirow{3}{*}{Token} & Training & 750k&47.4M\\
	                         &                                & Evaluation & 198k &12.5M\\
	                         &                                & Test &201k &12.6M\\
	                         \addlinespace[0.08cm]
	                         & \multirow{3}{*}{Construct} & Training &750k & 48.9M\\
	             Android     &                                    & Evaluation & 99k & 6.4M\\
	                         &                                    & Test & 101k &6.5M\\
	                         \addlinespace[0.08cm]
	                         & \multirow{3}{*}{Block} & Training & 205k &13.4M\\
	                         &                                & Evaluation & 27k &1.7M\\
	                         &                                & Test & 27k & 1.8M\\
\bottomrule
\end{tabular}
\vspace{-0.2cm} 
\end{table}

%% file: tables/hyperparameters.tex


\begin{table}[ht]
	\centering
	\caption{Hyperparameters Tuned for the \roberta Models.\vspace{-0.2cm}}
	\scriptsize
	\label{tab:hyper}
	\begin{tabular}{lll}
	\toprule
	{\bf Hyperparameter} & {\bf Experimented Values} & {\bf Best}\\\midrule
	Learning rate & \{$5e^{-5}$, $3e^{-5}$, $2e^{-5}$\} & $5e^{-5}$\\
	Batch size & \{16, 32, 64\} & 64\\
	\# Hidden Layers & \{6, 12, 16\} & 12\\
	\# Attention Heads & \{6, 12, 16\} & 16\\
	Hidden Layer Size & \{256, 512, 768, 1024\} & 768\\
	Intermediete Size & \{3072, 4096\} & 4,096\\
\bottomrule
\end{tabular} 
\vspace{-0.5cm}
\end{table}

%% file: results.tex
\section{Results Discussion} \label{sec:results}
\begin{figure*}[tb]
	\centering
	\includegraphics[width=0.9\linewidth]{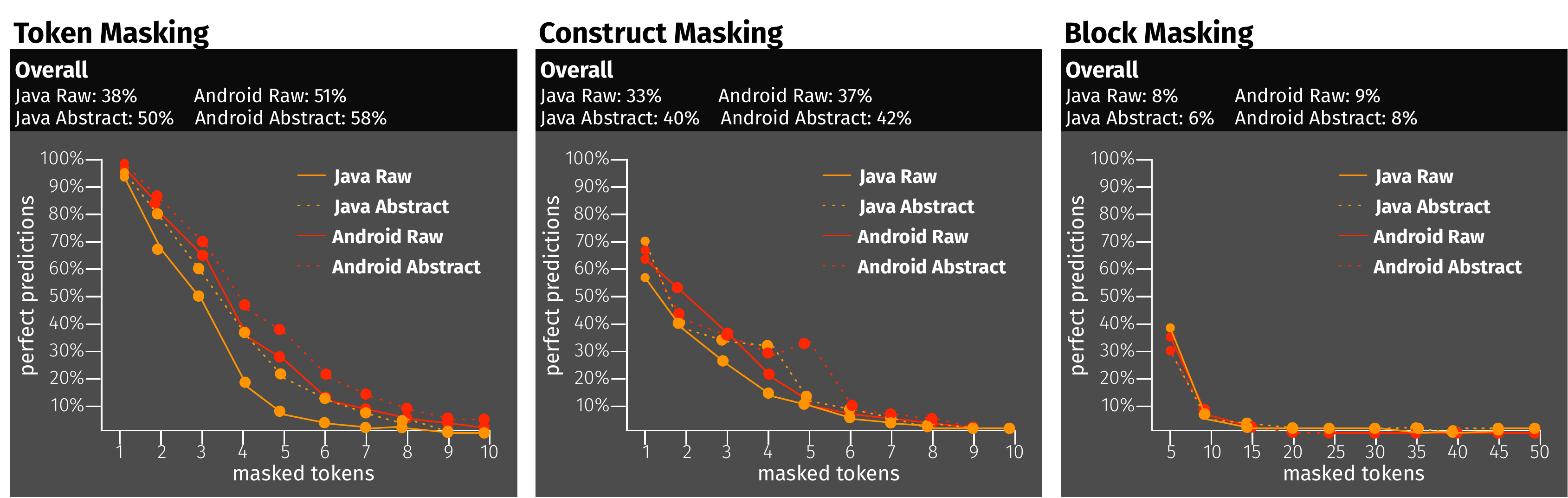}
	\caption{Results achieved by the \roberta model.\vspace{-0.4cm}}
	\label{fig:results}
\end{figure*}

\figref{fig:results} depicts the results achieved by the \roberta model in terms of perfect predictions in the different evaluation scenarios. Each sub-graph reports the results concerning a specific masking approach, namely (from left to right) \emph{token-masking}, \emph{construct-masking}, and \emph{block-masking}. The top part of each subgraph (\ie black background) shows the overall percentage of perfect predictions achieved on the Java and Android datasets when using the raw source code or its abstracted version. For example, 38\% of perfect predictions are generated by \roberta on the Java dataset when using token masking and the raw source code. 

The plots below show the percentage of perfect predictions by the number of masked tokens. For example, in the token masking scenario we randomly mask, for  each source code line $l$ having more than one token, its last $x$ tokens, where $x$ is a random number between $1$ $\dots$ $n-1$, with $n$ being the number of tokens of $l$, and $x$ is capped to a maximum of 10. 

The left graph in \figref{fig:results} shows the percentage of perfect predictions when we only mask the last token (\ie one masked token), the last two tokens, \etc The scale on the $x$ axis is different when dealing with the block masking scenario since here we mask entire blocks thus having, in some cases, dozens of masked tokens. Each point indicates that between $x-5$ and $x$ tokens were masked, \eg for the first data point at most 5 tokens were masked, for the second between 5 and 10, \etc

\tabref{tab:bleu} reports the average BLEU score in the four considered variants and the average normalized Levenshtein distance. Also in this case the results are grouped based on the masking level, dataset, and code representation.

\input{tables/bleu}

In the following, we summarize the quantitative results achieved. Then, we present qualitative examples of correct predictions made by the models and discuss the semantic equivalence of non-perfect predictions. Finally, we compare the performances of \roberta with that of the $n$-gram model. 

\subsection{Quantitative Results}

\textbf{Token masking.} The left part of \figref{fig:results} shows  that, as expected, the lower the number of masked tokens the higher the perfect predictions. Not surprisingly, the model is very effective when we only mask the last token in a statement. Indeed, in most cases, this will be a semicolon, a parenthesis, or a curly bracket. Thus, it is easy for the model to guess the last token. When moving to more challenging scenarios like the last five tokens masked in a statement, the percentage of perfect predictions is in the range $\sim$10-40\%, with major differences in the model effectiveness arising on the two considered datasets (Java and Android), and for the two code representations. As for the dataset, as we conjectured, the model achieves significantly better performance on the Android dataset (Fisher's test p-value$<$0.001), which is more specific and, thus, more subject to regularities in the source code. The gap in terms of perfect predictions is $\sim$20\% both when dealing with abstract and raw source code at $x=5$ (\ie last five tokens masked), with OR=1.35 and 1.69 respectively. For token masking, the red lines (Android) are always above the orange ones (Java), confirming the superior performance on the Android dataset (a more specific one) as a general trend.

As expected, the abstracted dataset leads to significantly better performances for both Java and Android datasets (McNemar test $p$-value$<$0.001, OR=3.66 and 1.96 respectively). This is due to the smaller vocabulary ensured by the abstraction and the simplification of the prediction task.

Looking at \tabref{tab:bleu}, the BLEU scores and the Levenshtein distance confirm what observed for perfect predictions: Performances for the Android dataset are better than for the Java one, and abstracted code outperforms raw code. All differences are statistically significant. For Android, 20\% and 24\% of predicted tokens must be changed with the abstract and raw representation, respectively, to obtain the reference code. Such a percentage grows for Java to 25\% (abstract) and 35\% (raw).

\textbf{Construct masking.} In this scenario (see central sub-graph in \figref{fig:results}), \roberta achieves above 50\% of perfect predictions when a single token is masked for both datasets/code representations.  Note that, in this scenario, also a single-token prediction is not trivial since we are in a context in which such a single token represents (i) the complete condition of an {\tt if} statement or a {\tt while}/{\tt for} loop, or (ii) the parameters in a method call, or (iii) the exception caught in a {\tt catch} statement. When the prediction is represented by a single token, it is usually related to a Boolean used in an {\tt if} condition (\eg {\tt if(true)}, {\tt if(valid)}, \etc) or the single parameter needed for a method invocation. 

Also in this case, a higher number of masked tokens implies lower performance, and the \roberta model confirms significantly better performance for the Android dataset although the gap is smaller (OR=1.19 for the raw and 1.12 for the abstract dataset). Again, code abstraction significantly helps the prediction (OR=1.72 for Java and 1.46 for Android). 

In terms of BLEU score and Levenshtein distance, the achieved values are worse as compared to the token-level masking, confirming the more challenging prediction scenario represented by the construct-level masking. On average, the developer may need to modify $\sim$40\% of the predicted tokens to obtain the reference code (small variations are observed among Java/Android and raw/abstract code). 

\textbf{Block masking.} This represents the most challenging prediction scenario for \roberta: The masked part can involve an entire statement or even span over two statements (maximum boundary we set). The performance of \roberta in terms of perfect predictions are above 30\% when dealing with small masked blocks, up to five tokens. These blocks are mostly related to \texttt{return} statements representing a code block (\eg the value to return when an \texttt{if} condition is satisfied), such as \texttt{ \{ return false; \}}, \texttt{ \{ return null; \}}, \etc

For longer  blocks, the performances  substantially drop. When considering blocks having between six and ten masked tokens, the percentage of perfect predictions is around 10\% for both datasets and code representations. The largest masked block reporting a perfect prediction is composed of 13 tokens for the Android raw and abstract datasets, 15 for the Java raw dataset, and 11 for the Java abstract dataset. 

Thus, at least with the amount of training we performed and the model architecture we used, \roberta is only able to correctly predict ``small'' masked blocks with good accuracy.  When masking code blocks, the performances of the abstract dataset are worse (OR=0.60 for Java and 0.71 for Android), whereas results for Android are only slightly better than for Java (OR=1.18 for the abstract and 1.08 for the raw dataset).

As expected, the BLEU scores are the lowest in this scenario (\tabref{tab:bleu}), and the developer may need to revise, on average, $\sim$55\% of the predicted tokens, independently from the dataset of interest and the used code representation. 

\subsection{Qualitative Analysis}
\begin{figure}[tb]
	\centering
	\includegraphics[width=0.9\linewidth]{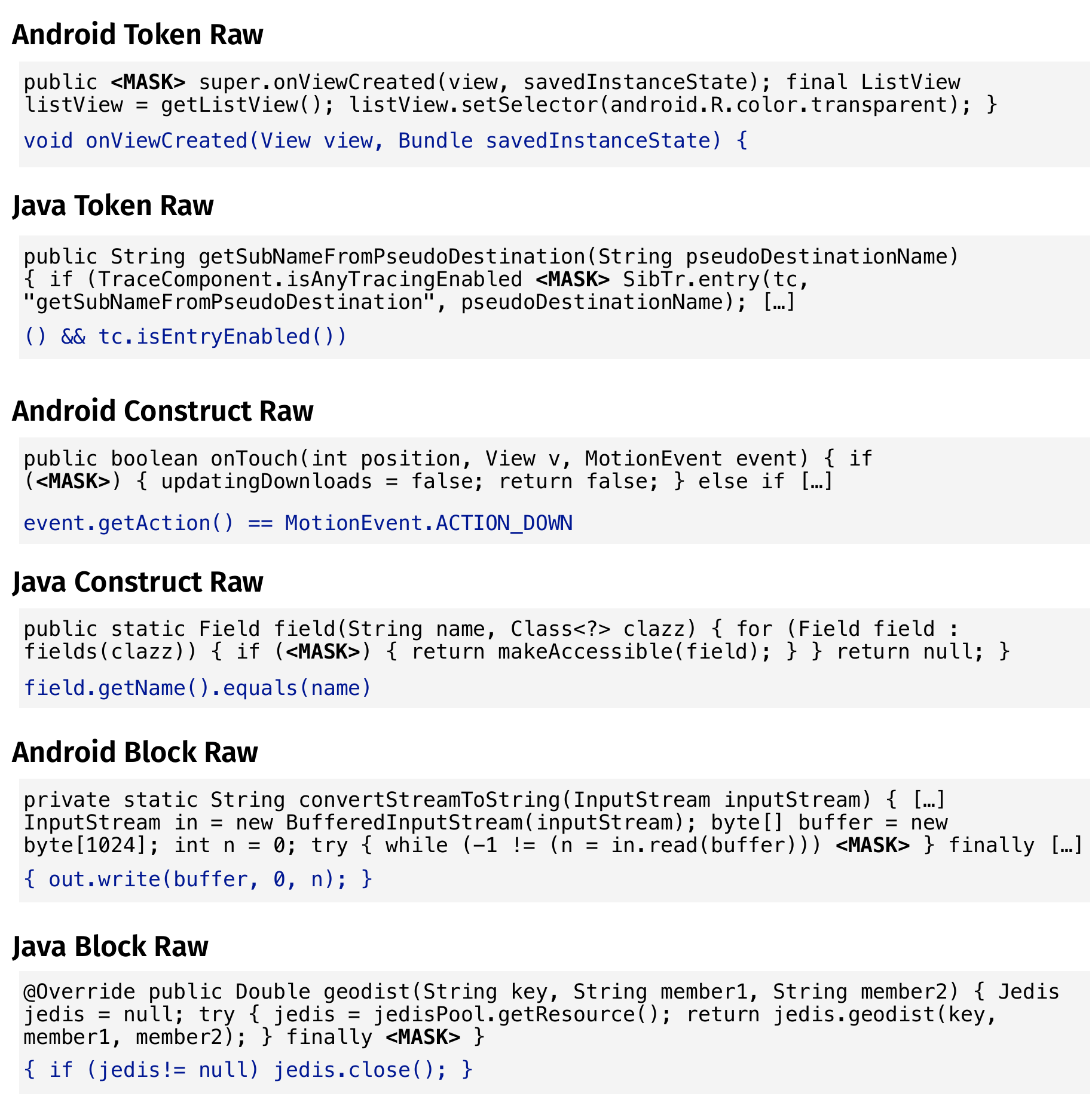}
	\caption{Examples of perfect predictions generated by \roberta\vspace{-0.6cm}}
	\label{fig:qualitative}
\end{figure}

\figref{fig:qualitative} reports qualitative examples of perfect predictions generated by \roberta. The black (top) code represents the masked code and the blue (bottom) one the code completion recommended by \roberta. Due to lack of space, we only report examples for the raw datasets. All predictions are available in our replication package \cite{replication}.

Besides this, as explained in \secref{sec:design}, we also manually analyzed 100 non-perfect predictions in each test dataset, to understand how many of them could be considered semantically equivalent to the reference code. We only found 35 out of the 1,200 cases in which the prediction could be considered as semantically equivalent. For example, the reference code was \texttt{return bitmap;} but \roberta predicted \texttt{\{return bitmap;\}}. Thus, we can estimate a 3\% of non-perfect predictions that could be counted as perfect ones. Not surprisingly, 33 of them had a Levenshtein distance lower than 0.25. Basically, the performance estimation provided in our study through the counting of the perfect predictions should be considered as a precise assessment of the model performances. 

\subsection{Comparison with an $n$-gram Model}
\tabref{tab:quantResults} reports the comparison in terms of perfect predictions between the \roberta models and the $n$-gram model \cite{Hellendoorn:fse2017}. One important clarification is needed to properly interpret the results of \tabref{tab:quantResults}. Since the two models use different scripts to tokenize the code, we excluded from the test sets cases in which the tokens to predict (\ie the masked ones) are tokenized differently between the two approaches (\eg one identifies 4 tokens and the other one 5). 

This resulted in the exclusions of a few hundred instances from each test set and explains the slightly different performances reported for \roberta between \tabref{tab:quantResults} and \figref{fig:results}. \roberta achieves better performance in all experimented datasets/code representations, and McNemar's test always indicates significant differences, with ORs ranging between 1.90 (block masking for Android raw) and 18.87 (construct masking for Android abstract).
In the token masking scenario, the performance of the $n$-gram model are very good and, especially on Java raw, close to the ones of \roberta. When masking specific constructs, the gap in performance becomes stronger (see \tabref{tab:quantResults}) with a substantial gap especially when dealing with abstract code. Finally, in the block masking experiment, both techniques struggle to obtain a high percentage of perfect predictions. However, \roberta consistently achieves a few percentage points more than the $n$-gram model.

While \roberta showed superior performance, there are two important aspects to consider. First, the $n$-gram model allows for  faster training. We estimate three to five times less training time needed for the $n$-gram model as compared to \roberta. We do not report precise data since such a study would require executing the training many times on the same machine, and such an analysis is out of the scope of this work. Once trained both models can generate predictions in fractions of a second. Second, in our evaluation, the $n$-gram cache model can leverage information about other code components coming from the same project (\eg same file or package \cite{Hellendoorn:fse2017}) of the method in which the prediction is performed. This is one of the advantages of the cache model \cite{Hellendoorn:fse2017} and, in a real scenario, it should be possible to use this information  assuming that the method on which the prediction is performed is not the first one written in the whole system. 

\input{tables/baseline-comparison}

While our design ensures that \roberta and the $n$-gram model leverage the same training information, we also experimented with the $n$-gram cache model in a scenario where the code from the ``test project'' is available when generating a prediction. For a given method $m_t$ in the test set, we clone its repository and  check if the source code of $m_t$ in the latest system snapshot is exactly the same as in the test set. If this is the case, we run the prediction on $m_t$ providing the cloned repository as a test folder, in such a way that it is leveraged by the cache model (this is done through the implementation of Hellendoorn \etal \cite{Hellendoorn:fse2017}).
If the method changed, we discard it and move to the next one. Since such a process is very expensive, we collected 200 data points on each raw test set, and we compare the performance of the $n$-gram model when such additional information is provided (and not) on these 200 instances. \tabref{tab:ngram} reports the achieved results. As expected, the performances of the $n$-gram model increase thanks to the use of the information in the test project. On these same 200 data points, the performances of \roberta are always superior but in the case of Java token masking. 

\input{tables/ngram}

%% file: tables/bleu.tex


\begin{table}[h]
	\centering
	\caption{\roberta BLEU score and Levenshtein distance.\vspace{-0.2cm}}
	\scriptsize
	\label{tab:bleu}
	\begin{tabular}{lrrrrr}
	\toprule
	\multicolumn{6}{c}{{\bf Token masking}}\\\midrule
	& \multicolumn{2}{c}{{\bf Java}} & & \multicolumn{2}{c}{{\bf Android}}\\ \cline{2-3} \cline{5-6}
	& {\bf Raw} & {\bf Abstract} & & {\bf Raw} & {\bf Abstract}\\\hline
	BLEU-1 & 0.60 & 0.73 && 0.73 & 0.78\\
	BLEU-2 & 0.43 & 0.59 && 0.61 & 0.68\\
	BLEU-3 & 0.23 & 0.41 && 0.44 & 0.53\\
	BLEU-4 & 0.10 & 0.27 && 0.28 & 0.39\\
	Levenshtein & 0.35 & 0.25 && 0.24 & 0.20\\\midrule

	\multicolumn{6}{c}{{\bf Construct masking}}\\\midrule
	& \multicolumn{2}{c}{{\bf Java}} & & \multicolumn{2}{c}{{\bf Android}}\\ \cline{2-3} \cline{5-6}
	& {\bf Raw} & {\bf Abstract} & & {\bf Raw} & {\bf Abstract}\\\hline
	BLEU-1 & 0.51 & 0.53 && 0.57 & 0.55\\
	BLEU-2 & 0.34 & 0.34 && 0.43 & 0.40\\
	BLEU-3 & 0.24 & 0.28 && 0.33 & 0.34\\
	BLEU-4 & 0.14 &0.17 && 0.26 & 0.28\\
	Levenshtein & 0.48 & 0.45 && 0.41 & 0.42 \\\midrule
	
	\multicolumn{6}{c}{{\bf Block masking}}\\\midrule
	& \multicolumn{2}{c}{{\bf Java}} & & \multicolumn{2}{c}{{\bf Android}}\\ \cline{2-3} \cline{5-6}
	& {\bf Raw} & {\bf Abstract} & & {\bf Raw} & {\bf Abstract}\\\hline
	BLEU-1 & 0.44 & 0.44 && 0.44 & 0.42\\
	BLEU-2 & 0.32 & 0.31 && 0.31 & 0.29\\
	BLEU-3 & 0.21 & 0.20 && 0.21 & 0.19\\
	BLEU-4 & 0.13 & 0.11 && 0.13 & 0.11\\
	Levenshtein & 0.54 & 0.55 &&0.55 & 0.56\\\bottomrule
\end{tabular} 
\vspace{-0.3cm}
\end{table}

%% file: tables/baseline-comparison.tex


\begin{table}[ht]
	\centering
	\caption{\roberta \emph{vs} n-gram: Perfect Predictions.\vspace{-0.2cm}}
	\scriptsize
	\label{tab:quantResults}
	\begin{tabular}{lrrrrr}
	\toprule
	\multicolumn{6}{c}{{\bf Token masking}}\\\midrule
	& \multicolumn{2}{c}{{\bf Java}} & & \multicolumn{2}{c}{{\bf Android}}\\ \cline{2-3} \cline{5-6}
	& {\bf Raw} & {\bf Abstract} & & {\bf Raw} & {\bf Abstract}\\\hline
	\roberta & 38.9\% & 50.7\% & & 51.8\% & 58.2\% \\
	n-gram & 30.4\% & 35.1\% & & 35.3\% & 41.1\% \\\midrule

	\multicolumn{6}{c}{{\bf Construct masking}}\\\midrule
	& \multicolumn{2}{c}{{\bf Java}} & & \multicolumn{2}{c}{{\bf Android}}\\ \cline{2-3} \cline{5-6}
	& {\bf Raw} & {\bf Abstract} & & {\bf Raw} & {\bf Abstract}\\\hline
	\roberta & 33.4\% & 40.4\% && 37.7\% & 43.1\% \\
	n-gram & 12.5\% & 10.7\% && 17.6\% & 10.7\% \\\midrule

	\multicolumn{6}{c}{{\bf Block masking}}\\\midrule
	& \multicolumn{2}{c}{{\bf Java}} & & \multicolumn{2}{c}{{\bf Android}}\\ \cline{2-3} \cline{5-6}
	& {\bf Raw} & {\bf Abstract} & & {\bf Raw} & {\bf Abstract}\\\hline
	\roberta & 8.7\% & 6.8\% && 9.3\% & 8.0\% \\
	n-gram & 4.5\% & 3.6\% && 6.6\% & 5.4\% \\\bottomrule
\end{tabular}
\vspace{-0.3cm}
\end{table}

%% file: tables/ngram.tex


\begin{table}[ht]
	\centering
	\caption{Perfect predictions of $n$-gram model when providing the cloned repository as test folder.\vspace{-0.2cm}}
	\scriptsize
	\label{tab:ngram}
	\begin{tabular}{lrrr}
	\toprule
	\multirow{2}{*}{{\bf Dataset}} & \multicolumn{2}{c}{{\bf n-gram}} & \multirow{2}{*}{{\bf RoBERTa}}\\ \cline{2-3}
	& {\bf Without Cloning} & {\bf With Cloning}\\\midrule
	Android token masking & 34.9\% & 42.2\% & 50.9\%\\
	Java token masking &  32.5\% & 43.8\% & 42.2\%\\
	Android construct masking & 13.9\% & 22.0\% & 37.8\%\\
	Java construct masking & 14.5\% & 20.5\% & 38.0\%\\
	Android block masking & 8.9\% & 11.8\% & 13.0\%\\
	Java block masking & 5.2\% & 8.4\% & 8.5\%\\
\bottomrule
\end{tabular} 
\vspace{-0.2cm}
\end{table}

%% file: threats.tex
\section{Threats to Validity} \label{sec:threats}

Threats to \emph{construct validity} concern the relationship between theory and observation.
One threat, also discussed by Hellendoorn \etal \cite{HellendoornPGB19}, is related to how we simulate the extent to which code completion intervenes during development, \ie by masking source code elements. As explained in \secref{sec:datasetProcessing}, we consider different masking levels, not only to evaluate the amount of code completion that can be predicted but also to simulate different ways a developer writes source code, especially because we cannot assume this is done sequentially. However, we are aware that the considered masking levels cover a limited number of cases that may not completely reflect how developers write code.


Threats to \emph{internal validity} concern factors, internal to our study, that could influence its results. To this extent, an important factor that influences DL performance is the calibration of hyperparameters, which has been performed as detailed in \secref{sec:analysis}. We are aware that due to feasibility reasons we only calibrated the hyperparameters on the abstract Android dataset, hence it is possible that a more specific calibration for each dataset would produce better performances.


Threats to  \emph{external validity} are related to the generalizability of our findings. On the one hand, we have evaluated \roberta performances on two large datasets, a generic one, and a specific one. At the same time, we do not know whether the obtained results generalize to different domains than Android, and other programming languages than Java.
A further threat is that our study is limited to the \roberta model for DL and, as a baseline for \textit{n}-gram models, the one Hellendoorn and Devanbu~\cite{Hellendoorn:fse2017}. Both represent the current state-of-the-art, however, it would be desirable to investigate how alternative approaches would work for the different evaluation scenarios.

%% file: related.tex
 
\section{Related Work} \label{sec:related}

\input{tables/related-comparison}

We detail the literature related to approaches (partially) automating the writing of new code. Due to lack of space, we do not discuss recently proposed techniques for automating bug-fixing \cite{Tufano:tosem2019,Chen:2019,Bader:oopsla2019}, learning code changes \cite{Tufano:icse2019,brody2020neural}, as well as source code search engines that can be used to identify pieces of code for reuse \cite{Bajracharya2006,Reiss2009,Thummalapenta2007b,Thummalapenta2008,Grechanik2010,McMillan2012}. 


The Prospector tool by Mandelin \etal \cite{MandelinXBK05} pioneered the area of code completion approaches, and aimed at suggesting, within the IDE, variables or method calls from the user's code base. Prospector was then followed by improvements such as the InSynth tool by Gvero \etal \cite{GveroKKP13} which, given a type expected at a given point in the source code, searches for type-compatible expressions.
Other approaches focus on specific elements of API usage completion. The work from Zhang \etal~\cite{ZhangYZFZZO12} aims at recommending parameter usages, achieving 64\% of useful recommendations and 53\% of perfect ones.

Bruch \etal~\cite{Bruch:fse2009} introduced the intelligent code completion system, able to filter out from the list of candidate method calls recommended by the IDE those that are more relevant to the current working context. 
Their results show the capability to correctly predict up to 82\% of method calls actually needed by developers, and up to 72\% of those that are relevant to the current development context. 
The approach by Bruch \etal has been improved by Proksch \etal~\cite{ProkschLM15}, by adding further contextual information and by proposing a Pattern-based Bayesian Networks approach. 
Differently from the aforementioned approaches, we do not restrict code completion to method calls.

Robbes and Lanza~\cite{Robb2010a} used information extracted from the change history of software systems to support the code completion of method calls and class names. 


Asaduzzaman \etal~\cite{Asaduzzaman2014} proposed a technique named CSCC (Context Sensitive Code Completion). They collect code examples from software repositories and, for each method call, represent its context as a set of methods, keywords, class and interface names appearing within four lines of code. 

This contextual information is then used to filter out method call recommendations. CSCC outperforms previous approaches, achieving 86\% precision and 99\% recall.

Hindle \etal~\cite{Hindle:icse2012} pioneered the work on statistical language models applied to software. They conceived the idea of ``naturalness of source code'' and used n-gram models to create a language-agnostic algorithm that is able to predict the next token in a given statement. 

Raychev \etal~\cite{Raychev:pldi14} approach the code completion problem through statistical language models. They extract sequences of method calls from a large code base, and use this dataset to train a language model able to predict API calls. Their model achieves a 90\% accuracy in the top-3 recommendations.

Nguyen \etal~\cite{Nguyen:icse2012} proposed GraPacc, a context-sensitive code completion model trained on a database of API usage patterns. These patterns are then matched to a given code under development to support code completion. GraPacc achieves up to 95\% precision and 92\% recall. A similar approach was later on proposed by Niu \etal~\cite{niu2017api} for API completion in Android: Given an API method as a query, their approach recommends a set of relevant API usage patterns. They report a 18\% improvement of F-Measure when comparing to pattern extraction using frequent-sequence mining. 

Tu \etal~\cite{Tu:fse2014} introduced a cache component to exploit the ``localness of code'' in the n-gram model. Results show that since code is locally repetitive, localized information can be used to improve performance. The enhanced model outperforms standard n-gram models by up to 45\% in accuracy. 

Hellendoorn and Devanbu~\cite{Hellendoorn:fse2017} proposed further improvements to the cached models aimed at considering specific characteristics of code (\eg unlimited, nested, and scoped vocabulary). Then, they compare their model with DL-based models, showing its superiority. Also, they show that the two families of techniques can be combined together, leading to an unprecedented 1.25 bits of entropy per token. The findings of this study showed that DL, with the considered limitations, was not the best technique for modeling source code.

Karampatsis \etal~\cite{Karampatsis:DLareBest}, a few years later, suggested instead that neural networks are the best language-agnostic algorithm for code completion. They proposed to overcome the \emph{out of vocabulary problem} by using \textit{Byte Pair Encoding} \cite{bpe}. In addition, the proposed neural network is able to dynamically adapt to different projects. Their best model outperforms n-gram models, achieving an entropy of 1.03 bits. 

Kim \etal~\cite{kim2020code} leveraged the Transformers neural network architecture for code completion. They provide the syntactic structure of code to the network by using information from the Abstract Syntax Tree to fortify the self-attention mechanism. Among the several models they experiment with, the best one reached a MRR up to 74.1\% in predicting the next token.

Alon \etal~\cite{alon2019structural} addressed the problem of code completion with a language-agnostic approach named Structural Language Model. It leverages the syntax to model the code snippet as a tree. The model, based on LSTMs and Transformers, receives an AST representing a partial expression (statement), with some missing consecutive tokens to complete. 

Their best model reached state-of-the-art performance with an exact match accuracy for the top prediction of 18.04\%.

Svyatkovskiy \etal~\cite{svyatkovskiy2020intellicode} introduced IntelliCode Compose, a general-purpose multilingual code completion tool capable of predicting code sequences of arbitrary token types. They do not leverage high-level structural representation, such as AST, and use subtokens to overcome the \emph{out of vocabulary problem}. Their model can recommend an entire statement, and achieves a perplexity of 1.82 for Python programming language.

\revised{Liu \etal~\cite{Liu:ase2020} presented a Transformer-based neural architecture pre-trained with the goal of incorporating both code understanding and generation tasks. Afterwards, the model was then fine-tuned on the classic code completion task.}

A problem related to code completion has also been tackled by Watson \etal \cite{Watson:icse2020}: The authors exploit a sequence-to-sequence model to recommend assert statements for a given Java test case. This technique is able to generate a specific type of code statement, with a top-1 accuracy of 31\%. \revised{Also, Kanade \etal \cite{kanade2020} show how code embeddings can support code-related tasks, including  \emph{variable misuse and repair}, related to code completion when focusing on a single token.}

Svyatkovskiy \etal \cite{svyatkovskiy2020fast} proposed a different perspective on neural code completion, shifting from a generative task to a learning-to-rank task. Their model is used to rerank the recommendations provided via static analysis, being cheaper in terms of memory footprint as compared to generative models. 

In terms of other studies related to the applicability of code completion to practice, Hellendoorn \etal \cite{HellendoornPGB19} studied 15,000 real code completions from 66 developers and found that typically-used code completion benchmarks --- \eg produced by artificially masking tokens --- may misrepresent actual code completion tasks. The study by Hellendoorn \etal suggests that further research is needed to assess the actual applicability of DL-based code completion to the real-world. This is however out of scope for our work, because our aim is to assess the capability of DL models to predict non-trivial portions of code going beyond a single method call or parameter.

\tabref{tab:related} summarizes the most related works (\ie the ones related to DL generative models for code completion) and compares it to our work.  To the best of our knowledge, our work is the first to present a comprehensive study on the effectiveness of a BERT model for the task of code completion. Indeed, all the previous techniques/studies dealing with code completion are limited to the generation of missing tokens in a single statement, while \textit{we push this problem forward by attempting the automatic generation of an entire code block} (\eg the body of a {\tt for} statement). 


%% file: tables/related-comparison.tex

\begin{table*}[ht]
	\centering
	\caption{Summary of previous studies using deep learning for generative code completion.\vspace{-0.2cm}}
	\scriptsize
	\begin{tabular}{@{}p{28em}p{7em}p{13em}p{8em}p{9em}@{}}
		\toprule
		\textbf{Reference} & \textbf{Model}  & \textbf{Granularity} & \textbf{Training} & \textbf{Performance metrics} \\
		\midrule
		\textit{Alon \etal (2019)~\cite{alon2019structural}}  \newline
		The authors leverage the Transformers architecture to complete multiple contiguous masked tokens in a given statement. 
		& LSTM and Transformers & Multiple contiguous tokens in a statement & 1.3M Java examples \newline 16k C\# examples& Exact match accuracy\\ 
		\midrule
		\textit{Karampatsis \etal (2019)~\cite{Karampatsis:DLareBest}} \newline
		The authors overcome the \emph{out of vocabulary} problems when using  neural networks for token-level code completion through BPE. 
		& GRU Neural Language Model with BPE & Single token & Java (1.5B tokens) \newline C (1.7B tokens) \newline Python (1B tokens) &  Entropy and MRR \\ 
		\midrule
		\textit{Kim \etal (2020)~\cite{kim2020code}} \newline
		The authors leverage the Transformers architecture and exploit structure information of code to support token-level code completion.
		& Transformers & Single token & 100k Python2 source code files & MRR \\ 
		\midrule
		\textit{Liu \etal (2020)~\cite{Liu:ase2020}} \newline
		The authors propose a multi-task code completion, with pre-training plus fine tuning on token types.
		& Transformers with pre-training &  Single token &  Java (6.9M tokens) \newline Typescript (1.1M tokens) & Top-1 accuracy \\
		\midrule
		\textit{Svyatkovskiy \etal (2020)~\cite{svyatkovskiy2020intellicode}} \newline
		The author propose a general-purpose multilingual code completion tool based on Transformers and BPE. 
		 & Transformers with BPE & Multiple contiguous tokens in a statement & 1.2B lines of code &  Similarity and Perplexity \\ 
		\midrule

		\textbf{Our work} \newline
		Experiment with BERT language model to test its limits for code completion. & RoBERTa with BPE & Multiple contiguous tokens in a statement \newline Specific code constructs \newline Entire statement(s) & Java (446M tokens) & Exact match accuracy \newline BLEU score \newline Levenshtein distance \newline Manual Analysis\\
		\bottomrule
	\end{tabular}%
	\label{tab:related}%
\end{table*}

%% file: conclusion.tex
\section{Conclusion} \label{sec:conclusion}
We empirically evaluated the performances of \roberta models in the task of code completion. We considered several code completion scenarios, moving from a few code tokens masked to entire code blocks. In a nutshell, the achieved results showed that: (i) \roberta achieves promising results, superior to a state-of-the-art $n$-gram model \cite{Hellendoorn:fse2017}; (ii) the models learn better on more specific datasets (\eg on Android rather than Java) and when code abstraction is used; (iii) while effective when a limited number of code tokens is masked (up to ten), \roberta models suffer from more challenging code completion tasks involving a higher number of tokens.

Our future research agenda in pushing forward automatic code generation includes  experimenting with other DL-based architectures (\eg T5 \cite{t5}).

%% file: main.bbl
\begin{thebibliography}{10}
\providecommand{\url}[1]{#1}
\csname url@samestyle\endcsname
\providecommand{\newblock}{\relax}
\providecommand{\bibinfo}[2]{#2}
\providecommand{\BIBentrySTDinterwordspacing}{\spaceskip=0pt\relax}
\providecommand{\BIBentryALTinterwordstretchfactor}{4}
\providecommand{\BIBentryALTinterwordspacing}{\spaceskip=\fontdimen2\font plus
\BIBentryALTinterwordstretchfactor\fontdimen3\font minus
  \fontdimen4\font\relax}
\providecommand{\BIBforeignlanguage}[2]{{%
\expandafter\ifx\csname l@#1\endcsname\relax
\typeout{** WARNING: IEEEtran.bst: No hyphenation pattern has been}%
\typeout{** loaded for the language `#1'. Using the pattern for}%
\typeout{** the default language instead.}%
\else
\language=\csname l@#1\endcsname
\fi
#2}}
\providecommand{\BIBdecl}{\relax}
\BIBdecl

\bibitem{Bruch:fse2009}
M.~Bruch, M.~Monperrus, and M.~Mezini, ``Learning from examples to improve code
  completion systems,'' in \emph{Proceedings of the 7th Joint Meeting of the
  European Software Engineering Conference and the ACM SIGSOFT Symposium on The
  Foundations of Software Engineering}, ser. ESEC/FSE 2009, 2009, pp. 213--222.

\bibitem{Robb2010a}
R.~Robbes and M.~Lanza, ``Improving code completion with program history,''
  \emph{Automated Software Engineering}, vol.~17, no.~2, pp. 181--212, 2010.

\bibitem{kim2020code}
S.~Kim, J.~Zhao, Y.~Tian, and S.~Chandra, ``Code prediction by feeding trees to
  transformers,'' \emph{arXiv preprint arXiv:2003.13848}, 2020.

\bibitem{han2009code}
S.~Han, D.~R. Wallace, and R.~C. Miller, ``Code completion from abbreviated
  input,'' in \emph{2009 IEEE/ACM International Conference on Automated
  Software Engineering}.\hskip 1em plus 0.5em minus 0.4em\relax IEEE, 2009, pp.
  332--343.

\bibitem{han2011code}
------, ``Code completion of multiple keywords from abbreviated input,''
  \emph{Automated Software Engineering}, vol.~18, no. 3-4, pp. 363--398, 2011.

\bibitem{Hindle:icse2012}
A.~Hindle, E.~T. Barr, Z.~Su, M.~Gabel, and P.~Devanbu, ``On the naturalness of
  software,'' in \emph{Proceedings of the 34th International Conference on
  Software Engineering}, ser. ICSE 2012.\hskip 1em plus 0.5em minus 0.4em\relax
  IEEE Press, 2012, pp. 837--847.

\bibitem{Nguyen:icse2012}
A.~T. {Nguyen}, T.~T. {Nguyen}, H.~A. {Nguyen}, A.~{Tamrawi}, H.~V. {Nguyen},
  J.~{Al-Kofahi}, and T.~N. {Nguyen}, ``Graph-based pattern-oriented,
  context-sensitive source code completion,'' in \emph{2012 34th International
  Conference on Software Engineering (ICSE)}, 2012, pp. 69--79.

\bibitem{Tu:fse2014}
Z.~Tu, Z.~Su, and P.~Devanbu, ``On the localness of software,'' in
  \emph{Proceedings of the 22nd ACM SIGSOFT International Symposium on
  Foundations of Software Engineering}, ser. FSE 2014, 2014, pp. 269--280.

\bibitem{Asaduzzaman2014}
M.~{Asaduzzaman}, C.~K. {Roy}, K.~A. {Schneider}, and D.~{Hou},
  ``Context-sensitive code completion tool for better api usability,'' in
  \emph{2014 IEEE International Conference on Software Maintenance and
  Evolution}, 2014, pp. 621--624.

\bibitem{Nguyen:msr2016}
A.~T. {Nguyen}, H.~A. {Nguyen}, and T.~N. {Nguyen}, ``A large-scale study on
  repetitiveness, containment, and composability of routines in open-source
  projects,'' in \emph{Proceedings of the IEEE/ACM 13th Working Conference on
  Mining Software Repositories (MSR 2016)}, 2016, pp. 362--373.

\bibitem{niu2017api}
H.~Niu, I.~Keivanloo, and Y.~Zou, ``Api usage pattern recommendation for
  software development,'' \emph{Journal of Systems and Software}, vol. 129, pp.
  127--139, 2017.

\bibitem{Hellendoorn:fse2017}
V.~J. Hellendoorn and P.~Devanbu, ``Are deep neural networks the best choice
  for modeling source code?'' in \emph{Proceedings of the 2017 11th Joint
  Meeting on Foundations of Software Engineering}, ser. ESEC/FSE 2017, 2017, p.
  763?773.

\bibitem{Karampatsis:DLareBest}
\BIBentryALTinterwordspacing
R.~Karampatsis and C.~A. Sutton, ``Maybe deep neural networks are the best
  choice for modeling source code,'' \emph{CoRR}, vol. abs/1903.05734, 2019.
  [Online]. Available: \url{http://arxiv.org/abs/1903.05734}
\BIBentrySTDinterwordspacing

\bibitem{alon2019structural}
U.~Alon, R.~Sadaka, O.~Levy, and E.~Yahav, ``Structural language models of
  code,'' \emph{arXiv}, pp. arXiv--1910, 2019.

\bibitem{svyatkovskiy2020intellicode}
A.~Svyatkovskiy, S.~K. Deng, S.~Fu, and N.~Sundaresan, ``Intellicode compose:
  Code generation using transformer,'' \emph{arXiv preprint arXiv:2005.08025},
  2020.

\bibitem{White2015}
\BIBentryALTinterwordspacing
M.~White, C.~Vendome, M.~Linares-V\'{a}squez, and D.~Poshyvanyk, ``Toward deep
  learning software repositories,'' in \emph{Proceedings of the 12th Working
  Conference on Mining Software Repositories}, ser. MSR '15.\hskip 1em plus
  0.5em minus 0.4em\relax Piscataway, NJ, USA: IEEE Press, 2015, pp. 334--345.
  [Online]. Available: \url{http://dl.acm.org/citation.cfm?id=2820518.2820559}
\BIBentrySTDinterwordspacing

\bibitem{roberta}
\BIBentryALTinterwordspacing
Y.~Liu, M.~Ott, N.~Goyal, J.~Du, M.~Joshi, D.~Chen, O.~Levy, M.~Lewis,
  L.~Zettlemoyer, and V.~Stoyanov, ``Roberta: {A} robustly optimized {BERT}
  pretraining approach,'' \emph{CoRR}, vol. abs/1907.11692, 2019. [Online].
  Available: \url{http://arxiv.org/abs/1907.11692}
\BIBentrySTDinterwordspacing

\bibitem{Delvin:2019}
J.~Devlin, M.-W. Chang, K.~Lee, and K.~Toutanova, ``{BERT}: Pre-training of
  deep bidirectional transformers for language understanding,'' in
  \emph{Proceedings of the 2019 Conference of the North {A}merican Chapter of
  the Association for Computational Linguistics: Human Language Technologies,
  Volume 1 (Long and Short Papers)}.\hskip 1em plus 0.5em minus 0.4em\relax
  Association for Computational Linguistics, Jun. 2019, pp. 4171--4186.

\bibitem{dreyer:bleu}
\BIBentryALTinterwordspacing
M.~Dreyer and D.~Marcu, ``{H}y{TER}: Meaning-equivalent semantics for
  translation evaluation,'' in \emph{Proceedings of the 2012 Conference of the
  North {A}merican Chapter of the Association for Computational Linguistics:
  Human Language Technologies}.\hskip 1em plus 0.5em minus 0.4em\relax
  Montr{\'e}al, Canada: Association for Computational Linguistics, Jun. 2012,
  pp. 162--171. [Online]. Available:
  \url{https://www.aclweb.org/anthology/N12-1017}
\BIBentrySTDinterwordspacing

\bibitem{levenshtein1966}
V.~Levenshtein, ``{Binary Codes Capable of Correcting Deletions, Insertions and
  Reversals},'' \emph{Soviet Physics Doklady}, vol.~10, p. 707, 1966.

\bibitem{attention}
\BIBentryALTinterwordspacing
A.~Vaswani, N.~Shazeer, N.~Parmar, J.~Uszkoreit, L.~Jones, A.~N. Gomez, L.~u.
  Kaiser, and I.~Polosukhin, ``Attention is all you need,'' in \emph{Advances
  in Neural Information Processing Systems 30}, I.~Guyon, U.~V. Luxburg,
  S.~Bengio, H.~Wallach, R.~Fergus, S.~Vishwanathan, and R.~Garnett, Eds.\hskip
  1em plus 0.5em minus 0.4em\relax Curran Associates, Inc., 2017, pp.
  5998--6008. [Online]. Available:
  \url{http://papers.nips.cc/paper/7181-attention-is-all-you-need.pdf}
\BIBentrySTDinterwordspacing

\bibitem{t5}
C.~Raffel, N.~Shazeer, A.~Roberts, K.~Lee, S.~Narang, M.~Matena, Y.~Zhou,
  W.~Li, and P.~J. Liu, ``Exploring the limits of transfer learning with a
  unified text-to-text transformer,'' 2019.

\bibitem{Arivazhagan:2019}
\BIBentryALTinterwordspacing
N.~Arivazhagan, A.~Bapna, O.~Firat, D.~Lepikhin, M.~Johnson, M.~Krikun, M.~X.
  Chen, Y.~Cao, G.~F. Foster, C.~Cherry, W.~Macherey, Z.~Chen, and Y.~Wu,
  ``Massively multilingual neural machine translation in the wild: Findings and
  challenges,'' \emph{CoRR}, vol. abs/1907.05019, 2019. [Online]. Available:
  \url{http://arxiv.org/abs/1907.05019}
\BIBentrySTDinterwordspacing

\bibitem{Wolf2019HuggingFacesTS}
T.~Wolf, L.~Debut, V.~Sanh, J.~Chaumond, C.~Delangue, A.~Moi, P.~Cistac,
  T.~Rault, R.~Louf, M.~Funtowicz, and J.~Brew, ``Huggingface's transformers:
  State-of-the-art natural language processing,'' \emph{ArXiv}, vol.
  abs/1910.03771, 2019.

\bibitem{bpe}
P.~Gage, ``A new algorithm for data compression,'' \emph{C Users J.}, vol.~12,
  no.~2, p. 23?38, 1994.

\bibitem{tokenizers}
\emph{Hugging Face's Tokenizer Repositor},
  \url{https://github.com/huggingface/tokenizers}.

\bibitem{karampatsis2020big}
R.-M. Karampatsis, H.~Babii, R.~Robbes, C.~Sutton, and A.~Janes, ``Big code !=
  big vocabulary: Open-vocabulary models for source code,'' in
  \emph{Proceedings of the 42nd International Conference on Software
  Engineering, {ICSE} 2020}, 2020, p. To Appear.

\bibitem{Tufano:icse2019}
M.~Tufano, J.~Pantiuchina, C.~Watson, G.~Bavota, and D.~Poshyvanyk, ``On
  learning meaningful code changes via neural machine translation,'' in
  \emph{Proceedings of the 41st International Conference on Software
  Engineering, {ICSE} 2019, Montreal, QC, Canada, May 25-31, 2019}, 2019, pp.
  25--36.

\bibitem{Tufano:tosem2019}
M.~Tufano, C.~Watson, G.~Bavota, M.~{Di Penta}, M.~White, and D.~Poshyvanyk,
  ``An empirical study on learning bug-fixing patches in the wild via neural
  machine translation,'' \emph{{ACM} Trans. Softw. Eng. Methodol.}, vol.~28,
  no.~4, pp. 19:1--19:29, 2019.

\bibitem{Java:CodeSearchNet}
\BIBentryALTinterwordspacing
H.~Husain, H.~Wu, T.~Gazit, M.~Allamanis, and M.~Brockschmidt, ``Codesearchnet
  challenge: Evaluating the state of semantic code search,'' \emph{CoRR}, vol.
  abs/1909.09436, 2019. [Online]. Available:
  \url{http://arxiv.org/abs/1909.09436}
\BIBentrySTDinterwordspacing

\bibitem{Deduplication}
M.~Allamanis, \emph{CodeSearchNet Deduplication Algorithm},
  \url{https://github.com/github/CodeSearchNet/blob/master/src/dataextraction/dedup_split.py}.

\bibitem{Geiger:2018}
\BIBentryALTinterwordspacing
F.-X. Geiger, I.~Malavolta, L.~Pascarella, F.~Palomba, D.~D. Nucci, and
  A.~Bacchelli, ``A graph-based dataset of commit history of real-world android
  apps,'' in \emph{Proceedings of the 15th International Conference on Mining
  Software Repositories, {MSR}}.\hskip 1em plus 0.5em minus 0.4em\relax ACM,
  May 2018. [Online]. Available: \url{https://androidtimemachine.github.io}
\BIBentrySTDinterwordspacing

\bibitem{SrcML}
\emph{SrcML Website}, \url{https://www.srcml.org/}.

\bibitem{wandb}
\emph{Weights and Biases Website}, \url{https://www.wandb.com/}.

\bibitem{Gu:2016}
\BIBentryALTinterwordspacing
X.~Gu, H.~Zhang, D.~Zhang, and S.~Kim, ``Deep api learning,'' in
  \emph{Proceedings of the 2016 24th ACM SIGSOFT International Symposium on
  Foundations of Software Engineering}, ser. FSE 2016.\hskip 1em plus 0.5em
  minus 0.4em\relax New York, NY, USA: ACM, 2016, pp. 631--642. [Online].
  Available: \url{http://doi.acm.org.proxy.wm.edu/10.1145/2950290.2950334}
\BIBentrySTDinterwordspacing

\bibitem{Jiang:ASE'17}
S.~Jiang, A.~Armaly, and C.~McMillan, ``Automatically generating commit
  messages from diffs using neural machine translation,'' in \emph{2017 32nd
  {{IEEE}}/{{ACM International Conference}} on {{Automated Software
  Engineering}} ({{ASE}})}, ser. ASE'17, Oct. 2017, pp. 135--146, iSSN:.

\bibitem{Watson:icse2020}
C.~Watson, M.~Tufano, K.~Moran, G.~Bavota, and D.~Poshyvanyk, ``On learning
  meaningful assert statements for unit test cases,'' in \emph{Proceedings of
  the 42nd International Conference on Software Engineering, {ICSE} 2020},
  2020, p. To Appear.

\bibitem{mcnemar}
Q.~McNemar, ``Note on the sampling error of the difference between correlated
  proportions or percentages,'' \emph{Psychometrika}, vol.~12, no.~2, pp.
  153--157, 1947.

\bibitem{bh}
B.~Yoav and H.~Yosef, ``Controlling the false discovery rate: A practical and
  powerful approach to multiple testing,'' \emph{Journal of the Royal
  Statistical Society. Series B (Methodological)}, vol.~57, no.~1, pp.
  289--300, 1995.

\bibitem{ngram}
\emph{N-gram Cached Model}, \url{https://github.com/SLP-team/SLP-Core}.

\bibitem{replication}
``Replication package~\url{https://github.com/RoBERTaCode/roberta}.''

\bibitem{HellendoornPGB19}
V.~J. Hellendoorn, S.~Proksch, H.~C. Gall, and A.~Bacchelli, ``When code
  completion fails: a case study on real-world completions,'' in
  \emph{Proceedings of the 41st International Conference on Software
  Engineering, {ICSE} 2019, Montreal, QC, Canada, May 25-31, 2019}, 2019, pp.
  960--970.

\bibitem{Liu:ase2020}
F.~Liu, G.~Li, Y.~Zhao, and Z.~Jin, ``Multi-task learning based pre-trained
  language model for code completion,'' in \emph{Proceedings of the 35th
  IEEE/ACM International Conference on Automated Software Engineering}, ser.
  ASE 2020.\hskip 1em plus 0.5em minus 0.4em\relax Association for Computing
  Machinery, 2020.

\bibitem{Chen:2019}
\BIBentryALTinterwordspacing
Z.~Chen, S.~Kommrusch, M.~Tufano, L.~Pouchet, D.~Poshyvanyk, and M.~Monperrus,
  ``Sequencer: Sequence-to-sequence learning for end-to-end program repair,''
  \emph{CoRR}, 2019. [Online]. Available: \url{http://arxiv.org/abs/1901.01808}
\BIBentrySTDinterwordspacing

\bibitem{Bader:oopsla2019}
J.~Bader, A.~Scott, M.~Pradel, and S.~Chandra, ``Getafix: learning to fix bugs
  automatically,'' \emph{Proc. {ACM} Program. Lang.}, vol.~3, no. {OOPSLA}, pp.
  159:1--159:27, 2019.

\bibitem{brody2020neural}
S.~Brody, U.~Alon, and E.~Yahav, ``Neural edit completion,'' \emph{arXiv
  preprint arXiv:2005.13209}, 2020.

\bibitem{Bajracharya2006}
S.~Bajracharya, T.~Ngo, E.~Linstead, Y.~Dou, P.~Rigor, P.~Baldi, and C.~Lopes,
  ``Sourcerer: A search engine for open source code supporting structure-based
  search,'' in \emph{Companion to the 21st ACM SIGPLAN Symposium on
  Object-Oriented Programming Systems, Languages, and Applications}, ser.
  OOPSLA ’06.\hskip 1em plus 0.5em minus 0.4em\relax ACM, 2006, p. 681–682.

\bibitem{Reiss2009}
S.~P. Reiss, ``Semantics-based code search,'' in \emph{Proceedings of the 31st
  International Conference on Software Engineering}, ser. ICSE ’09.\hskip 1em
  plus 0.5em minus 0.4em\relax IEEE Computer Society, 2009, p. 243–253.

\bibitem{Thummalapenta2007b}
S.~Thummalapenta and T.~Xie, ``Parseweb: A programmer assistant for reusing
  open source code on the web,'' in \emph{Proceedings of the Twenty-Second
  IEEE/ACM International Conference on Automated Software Engineering}, ser.
  ASE ’07.\hskip 1em plus 0.5em minus 0.4em\relax Association for Computing
  Machinery, 2007, p. 204–213.

\bibitem{Thummalapenta2008}
------, ``Spotweb: Detecting framework hotspots and coldspots via mining open
  source code on the web,'' in \emph{2008 23rd IEEE/ACM International
  Conference on Automated Software Engineering}, 2008, pp. 327--336.

\bibitem{Grechanik2010}
M.~Grechanik, C.~Fu, Q.~Xie, C.~McMillan, D.~Poshyvanyk, and C.~Cumby, ``A
  search engine for finding highly relevant applications,'' in
  \emph{Proceedings of the 32nd ACM/IEEE International Conference on Software
  Engineering - Volume 1}, ser. ICSE ’10.\hskip 1em plus 0.5em minus
  0.4em\relax ACM, 2010, p. 475–484.

\bibitem{McMillan2012}
C.~McMillan, M.~Grechanik, D.~Poshyvanyk, C.~Fu, and Q.~Xie, ``Exemplar: A
  source code search engine for finding highly relevant applications,''
  \emph{IEEE Transactions on Software Engineering}, vol.~38, no.~5, pp.
  1069--1087, 2012.

\bibitem{MandelinXBK05}
D.~Mandelin, L.~Xu, R.~Bod{\'{\i}}k, and D.~Kimelman, ``Jungloid mining:
  helping to navigate the {API} jungle,'' in \emph{Proceedings of the {ACM}
  {SIGPLAN} 2005 Conference on Programming Language Design and Implementation,
  Chicago, IL, USA, June 12-15, 2005}, 2005, pp. 48--61.

\bibitem{GveroKKP13}
T.~Gvero, V.~Kuncak, I.~Kuraj, and R.~Piskac, ``Complete completion using types
  and weights,'' in \emph{{ACM} {SIGPLAN} Conference on Programming Language
  Design and Implementation, {PLDI} '13, Seattle, WA, USA, June 16-19, 2013},
  2013, pp. 27--38.

\bibitem{ZhangYZFZZO12}
C.~Zhang, J.~Yang, Y.~Zhang, J.~Fan, X.~Zhang, J.~Zhao, and P.~Ou, ``Automatic
  parameter recommendation for practical {API} usage,'' in \emph{34th
  International Conference on Software Engineering, {ICSE} 2012, June 2-9,
  2012, Zurich, Switzerland}, 2012, pp. 826--836.

\bibitem{ProkschLM15}
S.~Proksch, J.~Lerch, and M.~Mezini, ``Intelligent code completion with
  bayesian networks,'' \emph{{ACM} Trans. Softw. Eng. Methodol.}, vol.~25,
  no.~1, pp. 3:1--3:31, 2015.

\bibitem{Raychev:pldi14}
V.~Raychev, M.~Vechev, and E.~Yahav, ``Code completion with statistical
  language models,'' in \emph{Proceedings of the 35th ACM SIGPLAN Conference on
  Programming Language Design and Implementation}, ser. PLDI 2014, 2014, pp.
  419--428.

\bibitem{kanade2020}
A.~Kanade, P.~Maniatis, G.~Balakrishnan, and K.~Shi, ``Learning and evaluating
  contextual embedding of source code,'' 2020.

\bibitem{svyatkovskiy2020fast}
A.~Svyatkovskiy, S.~Lee, A.~Hadjitofi, M.~Riechert, J.~Franco, and
  M.~Allamanis, ``Fast and memory-efficient neural code completion,'' 2020.

\end{thebibliography}
